\documentclass{article}

\usepackage{PRIMEarxiv}

\usepackage[utf8]{inputenc} 
\usepackage[T1]{fontenc}    
\usepackage{hyperref}       
\usepackage{url}            
\usepackage{booktabs}       
\usepackage{amsfonts}       
\usepackage{nicefrac}       
\usepackage{microtype}      
\usepackage{lipsum}
\usepackage{fancyhdr}       
\usepackage{graphicx}       
\graphicspath{{media/}}     

\title{Can Artificial Intelligence Transform DevOps?}

\author{
  Mamdouh Alenezi, Mohammad Zarour, and Mohammad Akour \\
  Software Engineering and Disruptive Innovation (SEDI) \\
  College of Computer and Information Sciences \\
  Prince Sultan University \\
  Riyadh, Saudi Arabia\\
  \texttt{sedi@psu.edu.sa}
}

\begin{document}
\maketitle

\begin{abstract}
DevOps and Artificial Intelligence (AI) are interconnected with each other. DevOps is a business-driven approach to providing quickly delivered quality software, and AI is the technology that can be used in the system to enhance its functionality. So, DevOps teams can use AI to test, code, release, monitor, and improve the system. Through AI, the automation process delivered by DevOps could be improved efficiently. This study aims to explore how AI can transform DevOps. The research is useful in terms of facilitating software developers and businesses to assess the importance of AI in DevOps. The study has practical implications as it elaborates on how AI transforms DevOps and in what way it can support businesses in their business.
\end{abstract}

\keywords{Software Engineering \and DevOps \and Artifical Intelligence}

\section{Introduction}

DevOps and Artificial Intelligence (AI) are interconnected with each other. Trivedi \cite{trivedi} proclaimed that DevOps is a business-driven approach to providing software, and AI is the technology that can be used in the system to enhance its functionality. So, DevOps teams can use AI to test, code, release, monitor, and improve the system. Through AI, the automation process delivered by DevOps could be improved efficiently. Besides, Battina \cite{battina2020devops} articulated that AI and DevOps are interdependent for the better performance of information systems. The author propounded that AI is significant for DevOps to perform better organizational functions. However, businesses lose focus on harnessing AI and DevOps to develop organizational systems better. In addition, Columbus \cite{columbus} discussed the ten ways AI can accelerate DevOps. The ten ways include enhancing the productivity of DevOps in real-time, streamlining the requirement management system, providing an effective bug detection system, prioritizing the security testing and results, improving software quality, providing adept troubleshooting, removing anomalies that can alter DevOps teams, enhancing supply chain management, improving traceability, and creating integrated DevOps strategy \cite{maheshwari2019digital,senapathi2018devops}. Moreover, \cite{columbus} articulated that it is paramount to discuss the transformation of DevOps through AI because it can help businesses realize its importance before it becomes late for them. So, the research motivation is to achieve the aim of the study. The study aims to assess whether AI can transform DevOps. The research can facilitate businesses and software developers to know the significance of AI in DevOps. They can use the study findings to elaborate on how AI transforms DevOps and how businesses can adapt it for their success. Hence, AI and DevOps are interconnected and can provide comprehensive benefits to businesses. The research shed light on discussing how AI can transform DevOps as a primary motivation. 

\subsection{DevOps: A Brief Overview}

DevOps is the ability of an organization to deliver applications and services at high velocity. Chapman \cite{chapman2014introduction} articulated that DevOps combines engineering patterns and practices, culture, and tools that can enhance the organizational capability to deliver services with increased quality. Different practices such as continuous integration, continuous delivery, monitoring and logging and infrastructure as code have evolved during DevOps adaption. Jha and Khan \cite{jha2018review} propounded that the objectives of DevOps include building up a culture of coordinated efforts, quickening time to advertise client input, keeping up with quality standards, considering experimentation, mechanizing conventional operational improvement, and using stages to give turnkey scenarios. Initially, DevOps was considered helpful for big businesses, but now every business is using DevOps to benefit from automation. In addition, Akshaya et al. \cite{akshaya2015basic} posited that 'Dev' means all developers involved in the production stage and 'Ops' means operation team, including security professionals, administration, system engineers and others. So, DevOps is an activity where developers and operational bodies collaborate in each DevOps cycle, starting from the development to the production stage. Hence, DevOps combines engineering, cultural, and IT activities to automate business operations or product delivery. It involves developers and operations together to enhance the quality of output \cite{bass2015devops, gupta2019challenges}. 

\subsection{Artificial Intelligence (AI): A Brief Overview}

Artificial intelligence (AI) is the intelligence demonstrated by machines rather than humans. Ertel \cite{ertel2018introduction} proclaimed that AI, also known as machine intelligence, demonstrates learning and intelligence in solving problems. In the information technology field, AI can be considered an intelligent agent who perceives the environment intelligently and takes actions to maximize the chances of achieving goals. As per the author, AI has entered our daily lives and is used by big organizations, for example, IBM, Uber, Google, Facebook, and others, to enhance the automated systems, i.e., DevOps, through innovation and continuous improvement promises. In a similar context, Flasiński \cite{flasinski2016introduction} elucidated that artificial intelligence is a logic-based approach. The AI logic represents the knowledge of the world's agent, its goals, and situational analysis through logic. It informs the agent about the current scenario that help in effective decision making. It reduces the human effort in making the decision. Hence, artificial intelligence is a logic-based natural computational language that perceives the environment and solves problems with intelligence and learning. It facilitates monitoring, assessing, and evaluation of systems and provide solution options to agents, which reduces decision-making time drastically.

\section{Literature Review}

\subsection{How do AI and DevOps work together?}

Artificial intelligence and DevOps are interconnected with each other. Kinsbruner \cite{kinsbruner} articulated that DevOps is about accelerating the software development process to deliver value. It enhances the collaboration between operation and developers to deliver the enhanced quality service to the client. Traditional DevOps only allowed organizations to implement CD/CI pipelines, but contemporary DevOps, in collaboration with AI, improved the operations and automated the code reviews. AI/ML automated DevOps' code reviews and analysis process and lessened the burden on teams. It facilitates to get earlier detection results of code flaws, security issues, and code-related defects that can solve in time and enhances the end-user performance results. In a similar context, Matsui et al. \cite{matsui2021applying} and Surya \cite{surya2021ai} articulated that AI transforms DevOps by enhancing the application's quality of development, deployment, and monitoring. It adds value to the DevOps cycle through automated feedback and reviews. The tasks are completed in less time, which lessens the delivery time of the application and increases the end-user satisfaction. Additionally, Ciucu et al. \cite{ciucu2019innovative} propounded that AI integrated with TensorFlow (TF), Caffe, Apache, and Microsoft Cognitive Toolkit in the past few years. Such DevOps deals with environment setup and database management. The integration of AI enhanced the quality of DevOps processes quality. For example, in a container pipeline, the implication of AI automates the monitoring of API-Calls, debugging information, and process states. Hence, AI and DevOps can work together by transforming DevOps processing time and enhancing the quality. 

\subsection{DevOps and AI as new Standard of Businesses}

It has been demonstrated that DevOps and AI can work together to enhance the efficiency of businesses. Battina \cite{battina2020devops} articulated that AI and DevOps are changing every industry by bringing automation and enhanced problem-solving solutions. The use of AI in mobile robots increases DevOps and AI to convert manual jobs to automated ones. They become the new standard of businesses without which it is impossible to succeed in the market \cite{yarlagadda2018rpa}. Moreover, Columbus \cite{columbus} discussed the ten ways AI can accelerate DevOps to enhance business efficiency. The ten ways include enhancing the productivity of DevOps in real-time, streamlining the requirement management system, providing an effective bug detection system, prioritizing the security testing and results, improving software quality, providing adept troubleshooting, can remove anomalies that can alter DevOps teams, enhancing supply chain management, improve traceability, and create integrated DevOps strategy. Hence, DevOps and AI are the future of businesses. AI is bringing significant changes in DevOps that can help businesses to achieve a competitive advantage in the market.

\subsection{Benefits and Challenges}

Integrating AI and DevOps can create overwhelming benefits and brings some challenges. The benefits of AI include facilitating the DevOps team in testing, coding, releasing, and enhancing the quality of programs. It converts the manual coding and reviewing to an automated system that decreases application delivery time and enhances problem-solving skills \cite{battina2020devops}. Additionally, 40\% of businesses will use AI and DevOps by 2024 because AI benefits businesses through transforming DevOps (Gartner). The transformation includes alleviating operational complexities, streamlining communication, simplifying application monitoring, guiding the approach for effective application development, fostering resolutions, and improving software testing. However, Hurst \cite{hurst} articulated that AI can distract the engineering team from end goals, bringing challenges in achieving the results. The narrow task intelligence or weak AI can provide risks by limiting the decision-making to a narrow set of statistics. So, developers and operational teams also need to use their understanding in decision-making. Hence, AI can transform DevOps to create overwhelming benefits for organizations. However, dependency on AI-driven statistics can bring challenges for organizations that need human-driven intelligence to control during the DevOps cycle of application delivery.

\section{Research Methodology}

In consequence of the stated objective of the study, the following are the preferred research methods. A qualitative method is applied for an in-depth and broader analysis of AI transforming DevOps. This method is useful in studies that demand non-statistical analysis. However, the quantitative method, on the other hand, is not preferred as it largely relies on using numerical, statistical and computational techniques \cite{quinlan2019business}, which do not correlate or are required for the present study. In relation to the data sources and data collection process, secondary data sources are preferred, which are easily available and accessible online, including books, articles, journals, websites, etc. However, the use of primary data sources is not needed in this study as it focuses on retrieving information and data from a group of people or first-hand sources. Moreover, primary sources are lengthy, time-consuming and expensive, and it fits well usually with the quantitative method in which surveys as a design are used \cite{quinlan2019business}. Therefore, relating to the present study, secondary sources are significant as they are not expensive, neither time-consuming nor difficult to interpret. Moreover, the information collected through secondary sources can be easily molded to suit the objectives of the study, while in primary sources, this cannot happen as the researcher have less control over the data/information. Hence, this secondary source for data collection is suitable to rely on online websites. In order to analyze the collected data, review-based analysis is considered, which is appropriate for comprehensively reviewing literature related to the topic and using it to support the key findings of the study \cite{long2014empirical}. In relevance to the qualitative method and secondary sources, review-based analysis is well-match and therefore is significant to accomplishing the key research objectives.

\section{Results and Discussion}

The findings of this study have focused on AI and DevOps integration, its link with business standards, and benefits and challenges. It is noted that both AI and DevOps are reliant as DevOps is a business-centric approach for delivering software, and the technology of AI can be connected to the system to increase functionality. Using AI, it is noted that DevOps can be tested, coded, released and can efficiently monitor software. AI is found useful in improving automation, addressing issues by identifying them quickly, and improving collaboration \cite{trivedi}. The findings of Ertel \cite{ertel2018introduction} correlate with the results arguing that AI is capable of solving problems and can be considered as an able instrument through which timely actions can be taken to increase the prospects of accomplishing key objectives. 

Moreover, the results also highlight that AI can have an important role in increasing the efficiency of DevOps, as it can increase performance by allowing quick operation and development of cycles and giving a compelling experience to customers. There are different ways through which AI can transform and work together with DevOps. Limited unregulated access to data is identified to be one of the key challenges encountered by DevOps teams, and therefore in this reference, integrating with AI can aid in free data from its company silos for data aggregation. In addition, AI is found to collect data from different sources and categorize it for repeatable and consistent analysis \cite{trivedi}. Likewise, the study of Kinsbruner \cite{kinsbruner} somewhat discussed similar that AI automated DevOps' analysis process facilitates teams and helps address issues and increase performance. 

Similarly, the results showed that Machine learning (ML) models could be helpful in case of failure in areas/tools in DevOps, which can slow down and weaken the process. ML can support predicting error using data, and AI can predict signs and read patterns of failure, particularly when a fault is identified. In addition, AI is useful in seeing indicators that cannot be perceived by humans, and early notification and prediction can help teams to acknowledge and fix the problem prior to it can affect SDLC (software development life cycle) \cite{trivedi}. Ciucu et al. \cite{ciucu2019innovative} support that the integration of AI increases the quality of DevOps processes, and AI and DevOps can be integrated by transforming the processing time of DevOps and increasing the quality. ML can enhance the performance of DevOps by reducing inefficiencies in SDLC and automating recurrent tasks. In addition, by using AI, the teams of DevOps can check and test software more efficiently and quickly.

With respect to the standard of business, the results showed that with the inclusion of ML and AI, the organization/business could be transformed digitally. The integration of AI and ML with DevOps results in significant shifts in its development, as initially, its sets DevOps as key support for the digital transformation of the business. For the organizations relying on living data, using ML and AI with DevOps is proven to increase their value greatly in different aspects, such as from efficient workflow to the strengthening of security for the development of applications. Considering the involvement of ML and AI, the automation and manual configuration of security aspects is fixated upon reducing the administration misconfiguration and changes of faults. Improvisation can be made by reducing downtime, and gaps can be done by assessing risks. Using ML and AI, data can be made efficient, as well as decision-making and analysis. 

A previous study by Battina \cite{battina2020devops} also highlighted that AI, and DevOps' collaboration and integration can increase business efficiency, and these can significantly affect the industry through automation and increase problem-solving solutions. It has been found that AI can increase DevOps for enhancing the efficiency of the business as well as its productivity by focusing on streamlining the essentialities of the management system, providing a bug detection system, and prioritizing the result and security test. Moreover, AI is observed to bring considerable transformation in DevOps that can support businesses in terms of their efficiency and competitiveness. 

Regarding the benefits and challenges, the results show that AI is beneficial in DevOps in terms of increasing the security of software. It does by raising the speed of security tests. Besides, it is observed that software developers are confused and in a position of indecisiveness when choosing to work on time-consuming and necessary security tests. However, by using AI, the software developers can easily run security tests and, at the same time, complete their work on deadlines. This is possible because AI can significantly lower the risk vector ID times, raising the efficiency of false-positive identities. Moreover, it has an important role in terms of the management of ill-trained security professionals, which is now a major problem for organizations.

Moreover, AI can make it done by assisting cybersecurity professionals and creating the demand for different types of security data professionals that are able to work using AI technologies. The findings of Battina \cite{battina2020devops} also agree that integrating AI can help to transform DevOps by reducing operational complexities, streamlining communication, improving software testing, simplifying monitoring of applications, fostering resolutions, and alleviating operational issues. On the contrary, there exist challenges as well, and it is found that a high level of complexity is linked to monitoring and handling the DevOps environment. Besides, there is a challenge for DevOps teams to manage the level in the contemporary distributed and dynamic application environment. The DevOps team has to manage data in Exabyte, and hence it turns out to be complex for people to manage such large data and address customer concerns, as it is a time-consuming and challenging task. Recent literature by Hurst \cite{hurst} highlights the same and argues that it can be challenging for humans to manage data. Moreover, it is noted that relying on AI-driven statistics can be complex for companies that require human-driven intelligence to control the DevOps cycle of application delivery. 

\section{Recommendations}

The following are the key recommendation in terms of AI transforming DevOps:

It is recommended to rely on AI for efficient application progress as the use of AI with instruments such as Git can give visibility to deal with irregularities in even resource handling, long build time, and code volume. Moreover, Ops teams for early detection can take the help of ML and AI as it allows detection of issues through quick mitigation response enabling business continuity. Moreover, AI can be significant in business assessment as well and helpful for ensuring business continuity for companies and facilitating business development. DevOps teams should use ML for emergency addressing as it can help assess machine intelligence and has a significant role in managing quick alerts by facilitating the system to point out the anomaly, therefore, aiding in the filtering process of unexpected alerts making it highly effective. Moreover, ML can be used in the production cycle and is found useful in assessing resource utilization. 

Likewise, AI can help transform DevOps through improved data access; therefore, it is recommended that it integrate AI as it facilitates compiling data from different sources and organizing it simultaneously. Moreover, it can be helpful in analyzing and providing an overview of trends. Moreover, AI can help in security as well; by using ML and AI, security threats can not only be identified but also managed successfully. Besides, AI can be proven effective in terms of bug detection; therefore, its application is recommended in this reference.

Columbus \cite{columbus} discussed that bug detection instruments at Facebook predicted faults and recommended solutions that proved to be 80\% correct, with AI instruments adapting to automatically fix the bugs. It is noted that Semmle Code QL is found to be a prime AI-centric DevOps instrument, and the Ops team can rely on this tool for tracking risk in code and identifying logical variants in the codebase. To add further, Microsoft has already used Code QL for identifying risks and vulnerabilities. Experts of security in the Microsoft team used this instrument to search for variants of major issues, enabling them to locate and respond to major code issues and avert further occurrences. Moreover, AI needs to be used by DevOps teams to analyze the success of projects in terms of supply chain management. AI with ML algorithms can find the pattern and acquire insight about the data and, therefore, can significantly aid DevOps teams. 

\section{Conclusion}

The study aimed to explore how AI can transform DevOps. The research is useful in terms of facilitating software developers and businesses to assess the importance of AI in DevOps. The study has practical implications as it elaborates on how AI transforms DevOps and in what way it can support businesses in their business. A plethora of literature has been reviewed explaining AI and DevOps, its integration and its benefits and challenges. The research adopted a qualitative method, secondary data source and review-based analysis. It is noted that DevOps, in collaboration with AI, is useful in improving operations and automating code reviews. Moreover, the automated code review and analysis process are likely to reduce the burden on DevOps teams and support them in getting early detection outcomes of code errors, security problems and code-associated defects. 

It is found that AI is important in increasing DevOps efficiency through quick operation and development of cycle. Incorporating both ML and AI with DevOps leads to considerable transformation in its development, and is increases value and ensures an efficient flow of work. Moreover, AI can help software developers to run security tests and timely complete work before deadlines. Nevertheless, the challenge related to AI is about managing data in Exabyte, which is complex for software developers and DevOps to manage, as it is quite a challenging task. Besides, key recommendations are also offered to businesses and software developers that can help them transform DevOps, including efficient application, ensuring business continuity, managing quick alerts, improved data access, aid in security, effective in bug detection, and Code QL.

\bibliographystyle{unsrt}  
\bibliography{references}

\begin{thebibliography}{10}

\bibitem{trivedi}
Mayank Trivedi.
\newblock 12 ways ai is transforming devops.
\newblock \emph{TA Digital}. Accessed May 01, 2022 [Online].

\bibitem{battina2020devops}
Dhaya~Sindhu Battina.
\newblock Devops, a new approach to cloud development \& testing.
\newblock {\em International Journal of Emerging Technologies and Innovative
  Research}, pages 2349--5162, 2020.

\bibitem{columbus}
Louis Columbus.
\newblock 10 ways ai is accelerating devops.
\newblock \emph{Forbes}. Accessed May 01, 2022 [Online].

\bibitem{maheshwari2019digital}
Anup Maheshwari.
\newblock {\em Digital transformation: Building intelligent enterprises}.
\newblock John Wiley \& Sons, 2019.

\bibitem{senapathi2018devops}
Mali Senapathi, Jim Buchan, and Hady Osman.
\newblock Devops capabilities, practices, and challenges: Insights from a case
  study.
\newblock In {\em Proceedings of the 22nd International Conference on
  Evaluation and Assessment in Software Engineering 2018}, pages 57--67, 2018.

\bibitem{chapman2014introduction}
David Chapman.
\newblock Introduction to devops on aws.
\newblock {\em Amazon Web Services}, 2014.

\bibitem{jha2018review}
Pratibha Jha and Rizwan Khan.
\newblock A review paper on devops: Beginning and more to know.
\newblock {\em International Journal of Computer Applications}, 180(48):16--20,
  2018.

\bibitem{akshaya2015basic}
HL~Akshaya, J~Vidya, and K~Veena.
\newblock A basic introduction to devops tools.
\newblock {\em International Journal of Computer Science \& Information
  Technologies}, 6(3):05--06, 2015.

\bibitem{bass2015devops}
Len Bass, Ingo Weber, and Liming Zhu.
\newblock {\em DevOps: A software architect's perspective}.
\newblock Addison-Wesley Professional, 2015.

\bibitem{gupta2019challenges}
Rajeev~Kumar Gupta, Mekanathan Venkatachalapathy, and Ferose~Khan Jeberla.
\newblock Challenges in adopting continuous delivery and devops in a globally
  distributed product team: a case study of a healthcare organization.
\newblock In {\em 2019 ACM/IEEE 14th International Conference on Global
  Software Engineering (ICGSE)}, pages 30--34. IEEE, 2019.

\bibitem{ertel2018introduction}
Wolfgang Ertel.
\newblock {\em Introduction to artificial intelligence}.
\newblock Springer, 2018.

\bibitem{flasinski2016introduction}
Mariusz Flasi{\'n}ski.
\newblock {\em Introduction to artificial intelligence}.
\newblock Springer, 2016.

\bibitem{kinsbruner}
Eran Kinsbruner.
\newblock How artificial intelligence (ai) and machine learning are changing
  devops.
\newblock \emph{Enterprisersproject.com}. Accessed May 01, 2022 [Online].

\bibitem{matsui2021applying}
Beatriz Mayumi~Andrade Matsui and Denise~Hideko Goya.
\newblock Applying devops to machine learning processes: A systematic mapping.
\newblock In {\em Anais do XVIII Encontro Nacional de Intelig{\^e}ncia
  Artificial e Computacional}, pages 559--570. SBC, 2021.

\bibitem{surya2021ai}
Lakshmisri Surya.
\newblock Ai and devops in information technology and its future in the united
  states.
\newblock {\em International Journal of Creative Research Thoughts (IJCRT)},
  pages 2320--2882, 2021.

\bibitem{ciucu2019innovative}
R~Ciucu, FC~Adochiei, Ioana-Raluca Adochiei, F~Argatu, GC~Seri{\c{t}}an,
  B~Enache, S~Grigorescu, and Violeta~Vasilica Argatu.
\newblock Innovative devops for artificial intelligence.
\newblock {\em The Scientific Bulletin of Electrical Engineering Faculty},
  19(1):58--63, 2019.

\bibitem{yarlagadda2018rpa}
Ravi~Teja Yarlagadda.
\newblock The rpa and ai automation.
\newblock {\em International Journal of Creative Research Thoughts (IJCRT),
  ISSN}, pages 2320--2882, 2018.

\bibitem{hurst}
Aaron Hurst.
\newblock The pros and cons of ai and ml in devops.
\newblock \emph{Information Age}. Accessed May 01, 2022 [Online].

\bibitem{quinlan2019business}
Christina Quinlan, Barry Babin, Jon Carr, and Mitch Griffin.
\newblock {\em Business research methods}.
\newblock South Western Cengage, 2019.

\bibitem{long2014empirical}
Haiying Long.
\newblock An empirical review of research methodologies and methods in
  creativity studies (2003--2012).
\newblock {\em Creativity Research Journal}, 26(4):427--438, 2014.

\end{thebibliography}

\end{document}